\documentstyle[prd,preprint,tighten,aps,floats,eqsecnum,%
amssymb,amsfonts,newlfont,graphicx]{revtex}
\begin{document}
\draft
\preprint{
\begin{tabular}{r}
   DFPD 00/EP/47
\\ KIAS-P00066
\\ arXiv:hep-ph/0010009
\end{tabular}
}
\title{Large $\nu_\mu\to\nu_\tau$ and $\nu_e\to\nu_\tau$
transitions in short-baseline experiments?}
\author{Carlo Giunti}
\address{INFN, Sez. di Torino, and Dip. di Fisica Teorica,
Univ. di Torino, I--10125 Torino, Italy, and
\\
School of Physics, Korea Institute for Advanced Study,
Seoul 130-012, Korea}
\author{Marco Laveder}
\address{Dip. di Fisica ``G. Galilei'', Univ. di Padova,
and INFN, Sez. di Padova, I--35131 Padova, Italy}
\date{October 13, 2000}
\maketitle
\begin{abstract}
Considering four-neutrino schemes
of type 3+1,
we identify four small regions of the
neutrino mixing parameter space
compatible with all data.
Assuming
a small mixing between the sterile neutrino and the
isolated mass eigenstate
we show that large
$\nu_\mu\to\nu_\tau$
and
$\nu_e\to\nu_\tau$
transitions are predicted in short-baseline experiments
and
could be observed
in the near future in dedicated experiments.
We discuss also implications for solar, atmospheric
and long-baseline
neutrino experiments
and we present a formalism that allows to
describe
in 3+1 schemes
atmospheric neutrino oscillations,
long-baseline $\nu_\mu$
disappearance and $\nu_\mu\to\nu_\tau$ transitions
in matter.
\end{abstract}

\section{Introduction}
\label{intro}

Neutrino oscillation experiments
are powerful probes of neutrino masses
(see \cite{Bilenky-Pontecorvo-PR-78,%
Mikheev-Smirnov-uspekhi-87,Bilenky-Petcov-RMP-87,%
Kuo-Pantaleone-RMP-89,CWKim-book,BGG-review-98}).
At present three types of neutrino oscillation experiments
have obtained positive results:
solar and atmospheric experiments
and one short-baseline experiment.

All solar neutrino experiments
(Homestake \cite{Homestake-98},
Kamiokande \cite{Kamiokande-sun-96},
GALLEX \cite{GALLEX-99},
SAGE \cite{SAGE-99},
Super-Kamiokande
\cite{SK-sun})
have found a deficit in the flux of electron neutrinos on Earth
with respect to the Standard Solar Model prediction
\cite{BP98},
which constitutes an indication in favor of
oscillations of electron neutrinos into other states.
Although
no direct proof of these oscillations exist at present,
there is a convincing evidence that
the deficit of solar $\nu_e$'s
is due to neutrino physics,
of which neutrino oscillations is the simplest
and most natural phenomenon.
Hopefully,
the issue will be definitively settled
in a few years by the new generation of neutrino
oscillation experiments
(GNO \cite{GNO},
SNO \cite{SNO},
Borexino \cite{Borexino},
ICARUS \cite{ICARUS}
and others \cite{vonFeilitzsch-Nu2000}).

Several atmospheric neutrino experiments
(Kamiokande \cite{Kam-atm},
IMB \cite{IMB},
Super-Kamiokande \cite{SK-atm},
Soudan 2 \cite{Soudan2},
MACRO \cite{MACRO})
have found
an anomalous ratio of the events
generated by muon and electron neutrinos
and an anomalous angular dependence
of the events generated by muon neutrinos.
Although so far the oscillation pattern
has not been observed,
this is considered as an evidence in favor
of oscillations of muon neutrinos into
tau or sterile neutrinos.
Transitions of muon neutrinos into electron neutrinos
are strongly disfavored by
the bounds established by the
long-baseline $\bar\nu_e$ disappearance experiments
CHOOZ \cite{CHOOZ}
and
Palo Verde \cite{PaloVerde}
and by the fact that
Super-Kamiokande data do not show any anomalous
angular dependence of the events
generated by atmospheric electron neutrinos. 
The present data of the Super-Kamiokande
\cite{SK-atm}
and MACRO
\cite{MACRO}
disfavor pure transitions of atmospheric muon neutrinos
into sterile states.
There are good chances that the issue will be clarified
in a definite way
in the near future
by long-baseline experiments with muon neutrino beams
(K2K \cite{K2K},
MINOS \cite{MINOS},
OPERA \cite{OPERA},
ICARUS \cite{ICARUS})
and by new atmospheric neutrino experiments
(MONOLITH \cite{MONOLITH}
and others \cite{Geiser-Nu2000}).
The K2K experiment have already
obtained some preliminary indication
of a possible transition
of muon neutrinos into other states
\cite{K2K-00}.

The third evidence in favor
of neutrino oscillations
has been found in the short-baseline LSND experiment
\cite{LSND},
where
an excess of $e^+$ events have been observed.
If interpreted in terms of
$\bar\nu_\mu\to\bar\nu_e$
neutrino oscillations,
this excess corresponds to an oscillation probability of
$(2.5\pm0.6\pm0.4)\times10^{-3}$.
This result has not been confirmed by other experiments
(many other less sensitive short-baseline
experiments have not found any signal of neutrino oscillations;
the strongest bounds have been obtained in the
Bugey \cite{Bugey},
CDHS \cite{CDHS},
CCFR \cite{CCFR},
BNL-E776 \cite{BNL-E776},
KARMEN \cite{KARMEN},
CHORUS \cite{CHORUS},
NOMAD \cite{NOMAD}
experiments),
but it is very interesting and important,
because it is the only existing evidence
of neutrino flavor transition
from one state to another
and because
such transitions could be explored
with high accuracy in future short-baseline experiments.
In a few years the
MiniBooNE experiment \cite{MiniBooNE}
will check the LSND signal
in terms of neutrino oscillations.

It is well known
(see \cite{Giunti-JHU-99,Fogli-Lisi-Marrone-Scioscia-no3-99})
that the three indications in favor
of neutrino oscillations need at least three different
neutrino mass-squared differences
($\Delta{m}^2$'s),
arranged in the hierarchical order
$
\Delta{m}^2_{\mathrm{SUN}}
\ll
\Delta{m}^2_{\mathrm{ATM}}
\ll
\Delta{m}^2_{\mathrm{SBL}}
$.
This means that at least four massive neutrinos must exist.
Here we consider the minimal case of four massive neutrinos,
that has been considered recently by many authors
(see \cite{BGG-AB-96,Barger-variations-98,BGGS-AB-99,%
DGKK-99,Concha-foursolar-00,Concha-foursolar-update,%
Yasuda-fouratm-00,%
Peres-Smirnov-3+1,%
Lisi-Nu2000,Fogli-Lisi-Marrone-fouratm-00,%
Barger-Fate-2000,BGG-review-98,Giunti-JHU-99}
and references therein),
whose flavor basis is constituted
by the standard active neutrinos
$\nu_e$, $\nu_\mu$, $\nu_\tau$,
and a sterile neutrino $\nu_s$.
Figure~\ref{4schemes}
shows the six possible four-neutrino schemes
that can accommodate the observed hierarchy of
$\Delta{m}^2$'s.
These six schemes are divided in two classes:
3+1 and 2+2.
In the 3+1
schemes there is a group of three neutrino masses
separated from an isolated mass
by the LSND gap of the order of 1 eV,
such that the largest
mass-squared difference,
$\Delta{m}^2_{41}$
(where
$\Delta{m}^2_{kj} \equiv m_k^2 - m_j^2$
and
$m_k$,
$m_j$
are neutrino masses,
with
$k,j=1,\ldots,4$),
generates the oscillations observed in the LSND experiment.
In the 2+2 schemes
there are two couples of close mass eigenstates
separated by the LSND gap.
The numbering of the mass eigenvalues in Fig.~\ref{4schemes}
is conveniently chosen in order to have always
solar neutrino oscillations generated by
$\Delta{m}^2_{21} = \Delta{m}^2_{\mathrm{SUN}}$
and short-baseline (SBL) oscillations generated by
$
|\Delta{m}^2_{41}|
\simeq
|\Delta{m}^2_{42}|
=
\Delta{m}^2_{\mathrm{SBL}}
$
(we have also
$
|\Delta{m}^2_{41}|
\simeq
|\Delta{m}^2_{43}|
$
in 3+1 schemes and
$
|\Delta{m}^2_{41}|
\simeq
|\Delta{m}^2_{31}|
\simeq
|\Delta{m}^2_{32}|
$
in 2+2 schemes).
In 3+1 schemes
atmospheric neutrino oscillations are generated by
$
|\Delta{m}^2_{31}|
\simeq
|\Delta{m}^2_{32}|
=
\Delta{m}^2_{\mathrm{ATM}}
$,
whereas in 2+2 schemes
they are generated by
$
|\Delta{m}^2_{43}|
=
\Delta{m}^2_{\mathrm{ATM}}
$.

\begin{figure}[p!]
\begin{center}
\includegraphics[bb=89 331 540 763,width=\textwidth]{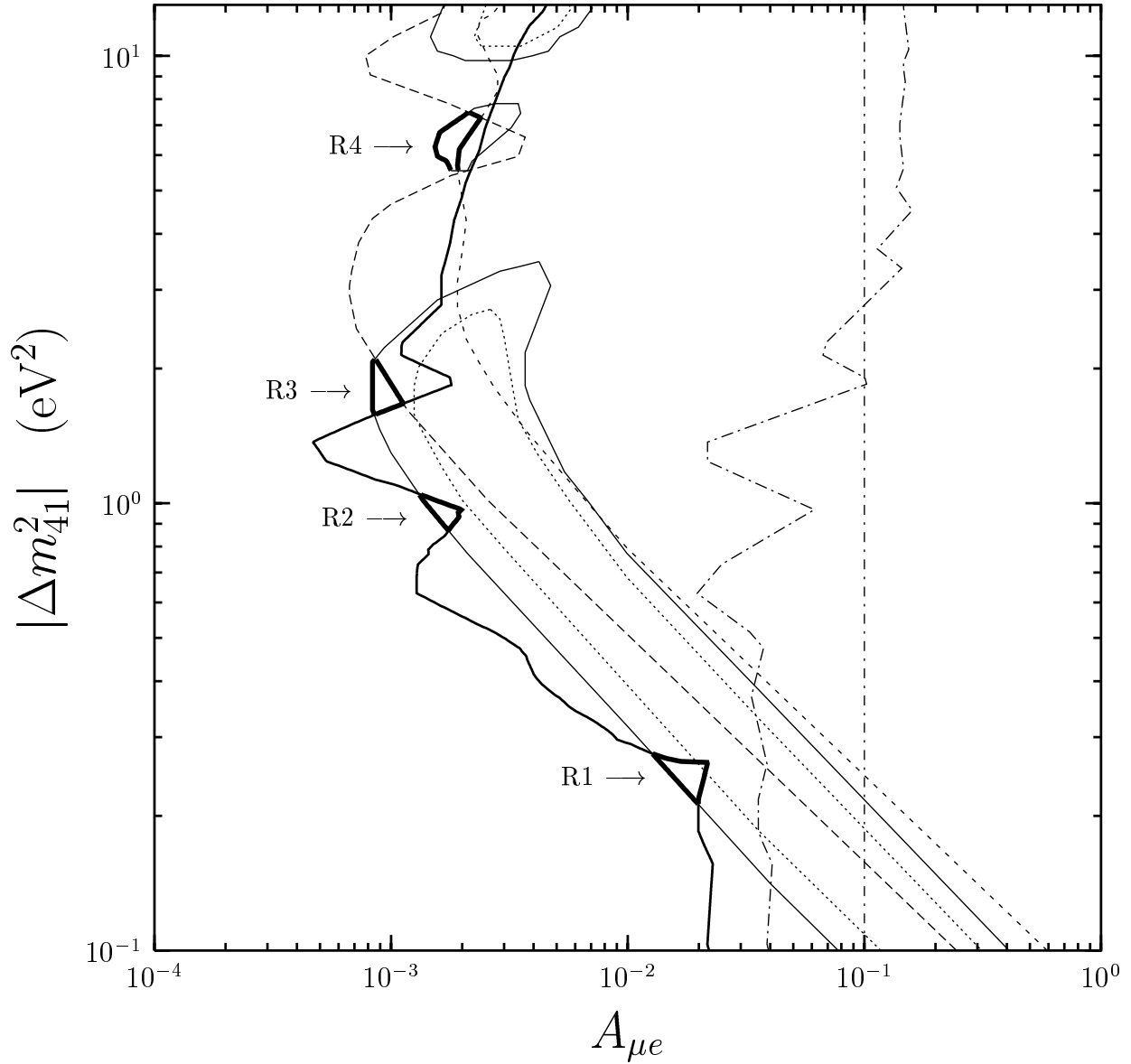}
\end{center}
\caption{ \label{amuel-allowed}
\textbf{Very Thick Solid Line:}
Allowed regions.
\textbf{Thick Solid Line:}
Disappearance bound (\ref{02}).
\textbf{Dotted Line:}
LSND 2000 allowed regions at 90\% CL \protect\cite{LSND}.
\textbf{Solid Line:}
LSND 2000 allowed regions at 99\% CL \protect\cite{LSND}.
\textbf{Broken Dash-Dotted Line:}
Bugey exclusion curve at 90\% CL \protect\cite{Bugey}.
\textbf{Vertical Dash-Dotted Line:}
CHOOZ exclusion curve at 90\% CL \protect\cite{CHOOZ}.
\textbf{Long-Dashed Line:}
KARMEN 2000 exclusion curve at 90\% CL \protect\cite{KARMEN}.
\textbf{Short-Dashed Line:}
BNL-E776 exclusion curve at 90\% CL \protect\cite{BNL-E776}.
}
\end{figure}

It has been shown that in the framework of four-neutrino mixing
the 3+1 schemes are disfavored by the experimental data,
with respect to the 2+2 schemes
\cite{BGG-AB-96,Barger-variations-98,BGGS-AB-99,Giunti-JHU-99}.
However,
in a recent paper
Barger, Kayser, Learned, Weiler and Whisnant
\cite{Barger-Fate-2000}
noticed that the new 99\% CL allowed region
obtained recently in the LSND experiment and presented at the
Neutrino 2000 conference
allows the existence 3+1 schemes
(see also \cite{Peres-Smirnov-3+1})
in four small regions of the amplitude
$A_{\mu e}$
of
$\nu_\mu\to\nu_e$ and $\bar\nu_\mu\to\bar\nu_e$
oscillations
($A_{\mu e}$
is equivalent to the usual
$\sin^2 2\vartheta$
in the two-generation case).
These regions,
to be derived later (Eqs.~(\ref{24})--(\ref{03}))
are shown in Fig.~\ref{amuel-allowed},
enclosed by very thick solid lines.
They lay at
\begin{eqnarray}
&&
\mbox{R1:}
\quad
|\Delta{m}^2_{41}| \simeq 0.25 \, \mathrm{eV}^2
\,,
\nonumber
\\
&&
\mbox{R2:}
\quad
|\Delta{m}^2_{41}| \simeq 0.9 \, \mathrm{eV}^2
\,,
\nonumber
\\
&&
\mbox{R3:}
\quad
|\Delta{m}^2_{41}| \simeq 1.7 \, \mathrm{eV}^2
\,,
\nonumber
\\
&&
\mbox{R4:}
\quad
|\Delta{m}^2_{41}| \simeq 6 \, \mathrm{eV}^2
\,.
\label{Regions}
\end{eqnarray}

Barger, Kayser, Learned, Weiler and Whisnant
explored the phenomenological consequences
of the assumption
\begin{equation}
1 - |U_{s4}|^2 \ll 1
\label{12}
\end{equation}
(here $U_{\alpha k}$
are the elements of the $4\times4$
neutrino mixing matrix
with
$\alpha=s,e,\mu,\tau$
and
$k=1,\ldots,4$).
Such a scheme is attractive because
it represents a perturbation of the standard
three-neutrino mixing
in which a mass eigenstate is added,
that mixes mainly with the new sterile neutrino $\nu_s$
and very weakly with
the standard active neutrinos $\nu_e$, $\nu_\mu$ and $\nu_\tau$.
In this case,
the usual phenomenology of three-neutrino mixing
in solar and atmospheric
neutrino oscillation experiments is practically unchanged.
The atmospheric neutrino anomaly would be explained by
dominant $\nu_\mu\to\nu_\tau$ transitions,
with possible sub-dominant $\nu_\mu\leftrightarrows\nu_e$
transitions constrained by the CHOOZ bound
\cite{Fogli-Lisi-Marrone-Montanino-subdominant-97,%
Bilenky-Giunti-CHOOZ-98,%
Lisi-Nu2000}.
The solar neutrino problem would be explained by an
approximately equal mixture of
$\nu_e\to\nu_\mu$ and $\nu_e\to\nu_\tau$ transitions
\cite{Bilenky-Giunti-CHOOZ-98,%
Fogli-Lisi-Montanino-Palazzo-sun-3nu-00}.

Here,
we consider another possibility that,
as we will see,
predicts relative large
$\nu_\mu\to\nu_\tau$ and $\nu_e\to\nu_\tau$ transitions
in short-baseline neutrino oscillation experiments,
that could be observed in the near future.
We consider the 3+1 schemes with
\begin{equation}
|U_{s4}|^2 \ll 1
\,.
\label{13}
\end{equation}
This could be obtained,
for example,
in the hierarchical scheme I (see Fig.~\ref{4schemes})
with an appropriate symmetry keeping the
sterile neutrino very light,
\textit{i.e.}
mostly mixed with the lightest mass eigenstates.
Notice that nothing forbids
$|U_{s4}|^2$
to be even zero exactly.

\section{3+1 schemes with $|U_{\lowercase{s4}}|^2\ll1$}
\label{small}

From the assumption (\ref{13}),
since the amplitude of
$\nu_\alpha\to\nu_\beta$ and $\nu_\beta\to\nu_\alpha$
oscillations in short-baseline neutrino oscillation experiments
(equivalent to the usual
$\sin^2 2\vartheta$
in the two-generation case)
is given by
\cite{BGG-AB-96,BGG-review-98}
\begin{equation}
A_{\alpha\beta}
=
A_{\beta\alpha}
=
4 |U_{\alpha4}|^2 |U_{\beta4}|^2
\qquad
(\alpha,\beta=s,e,\mu,\tau)
\,,
\label{14}
\end{equation}
we have
\begin{equation}
A_{{\alpha}s}
\ll
1
\qquad
(\alpha=e,\mu,\tau)
\,,
\label{15}
\end{equation}
\textit{i.e.}
the transitions
from active to sterile neutrinos
in short-baseline experiments are strongly suppressed.
In the following we will neglect them.

Let us consider now the oscillation amplitude
(equivalent to the usual two-generation $\sin^2 2\vartheta$)
in short-baseline $\nu_\alpha$ disappearance experiments,
which is given by
\cite{BGG-AB-96,BGG-review-98}
\begin{equation}
B_\alpha
=
|U_{\alpha4}|^2 \left( 1 - |U_{\alpha4}|^2 \right)
\,.
\label{16}
\end{equation}
In general the oscillation amplitude 
in $\nu_\alpha$ disappearance experiments
is related to the amplitude of
$\nu_\alpha\to\nu_\beta$
oscillations by the relation
\begin{equation}
B_\alpha
=
\sum_{\beta\neq\alpha} A_{\alpha\beta}
\,,
\label{17}
\end{equation}
that quantify the conservation of probability.
Using Eq.~(\ref{15}),
Eq.~(\ref{17}) gives
\begin{eqnarray}
&&
A_{e\tau} \simeq B_e - A_{{\mu}e}
\,.
\label{18}
\\
&&
A_{\mu\tau} \simeq B_\mu - A_{{\mu}e}
\,.
\label{19}
\end{eqnarray}
Therefore,
we have
\begin{eqnarray}
&&
B_e^{\mathrm{min}} - A_{{\mu}e}^{\mathrm{max}}
\lesssim
A_{e\tau}
\lesssim
B_e^{\mathrm{max}} - A_{{\mu}e}^{\mathrm{min}}
\,.
\label{21}
\\
&&
B_\mu^{\mathrm{min}} - A_{{\mu}e}^{\mathrm{max}}
\lesssim
A_{\mu\tau}
\lesssim
B_\mu^{\mathrm{max}} - A_{{\mu}e}^{\mathrm{min}}
\,.
\label{22}
\end{eqnarray}
Let us determine
$B_e^{\mathrm{min}}$,
$B_e^{\mathrm{max}}$,
$B_\mu^{\mathrm{min}}$,
$B_\mu^{\mathrm{max}}$,
$A_{{\mu}e}^{\mathrm{min}}$,
$A_{{\mu}e}^{\mathrm{max}}$
from the results of short-baseline experiments.

\section{General bounds in 3+1 schemes
and $\nu_\mu\to\nu_{\lowercase{e}}$ short-baseline transitions}
\label{General}

The values of
$B_e^{\mathrm{max}}$
and
$B_\mu^{\mathrm{max}}$
are given by
the exclusion plots of $\bar\nu_e$ and $\nu_\mu$
disappearance experiments
(notice that
$B_e^{\mathrm{max}}$
and
$B_\mu^{\mathrm{max}}$
depend on
$|\Delta{m}^2_{41}|$).
The most stringent bounds
for
$|\Delta{m}^2_{41}|$
in the LSND-allowed region
are given by
the exclusion curves of the Bugey \cite{Bugey}
and CHOOZ \cite{CHOOZ}
reactor $\bar\nu_e$
disappearance experiments
and
the exclusion curve
of the CDHS accelerator $\nu_\mu$ disappearance experiment
\cite{CDHS}.

\begin{figure}[p!]
\begin{center}
\includegraphics[bb=89 331 540 763,width=\textwidth]{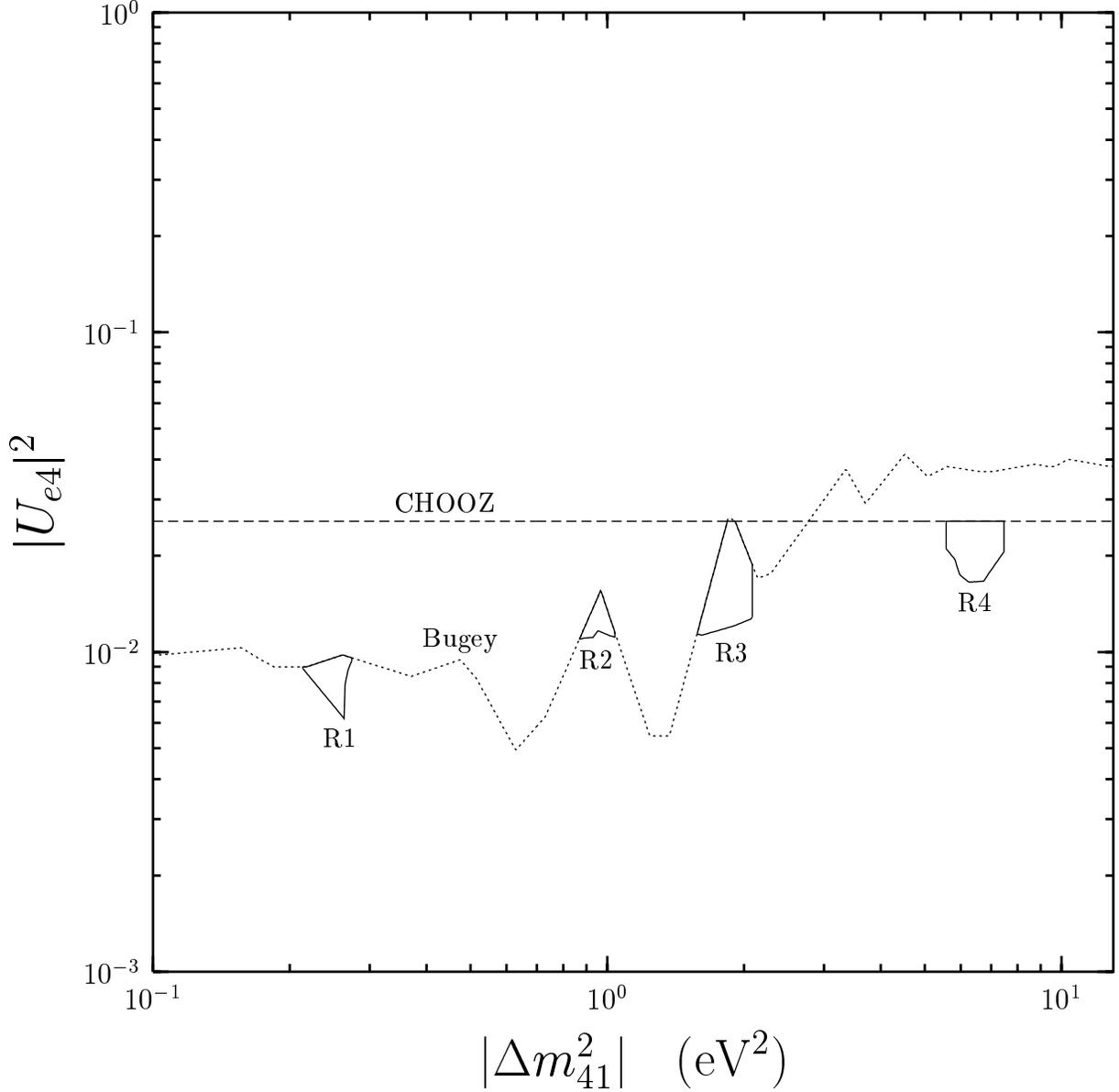}
\end{center}
\caption{ \label{uel4}
\textbf{Dotted Line:}
$|U_{e4}|^2_{\mathrm{max}}$
obtained from
the 90\% CL exclusion curve of the
Bugey short-baseline
reactor $\bar\nu_e$ disappearance experiment
\protect\cite{Bugey}
(Eq.~(\ref{25})).
\textbf{Dashed Line:}
$|U_{e4}|^2_{\mathrm{max}}$
obtained from
the 90\% CL exclusion curve of the
CHOOZ long-baseline
reactor $\bar\nu_e$ disappearance experiment
\protect\cite{CHOOZ}
(Eq.~(\ref{25})).
\textbf{Solid Line:}
Allowed regions (Eq.~(\ref{06})).
}
\end{figure}

The bounds
$B_e \leq B_e^{\mathrm{max}}$
and
$B_\mu \leq B_\mu^{\mathrm{max}}$,
together with the results of solar and atmospheric
neutrino experiments
imply that
$|U_{e4}|^2$
and
$|U_{\mu4}|^2$
are small
\cite{BGG-AB-96,BGGS-AB-99,Giunti-JHU-99}:
\begin{equation}
|U_{e4}|^2 \leq |U_{e4}|^2_{\mathrm{max}}
\quad \mbox{and} \quad
|U_{\mu4}|^2 \leq |U_{\mu4}|^2_{\mathrm{max}}
\,,
\label{24}
\end{equation}
where
$|U_{e4}|^2_{\mathrm{max}}$ and $|U_{\mu4}|^2_{\mathrm{max}}$
are given by
\begin{eqnarray}
&&
|U_{e4}|^2_{\mathrm{max}}
=
\frac{1}{2}
\left( 1 - \sqrt{ 1 - B_e^{\mathrm{max}} } \, \right)
\,,
\label{25}
\\
&&
|U_{\mu4}|^2_{\mathrm{max}}
=
\min\left[
\frac{1}{2}
\left( 1 - \sqrt{ 1 - B_\mu^{\mathrm{max}} } \, \right)
\, , \,
0.55
\right]
\,.
\label{26}
\end{eqnarray}
The number 0.55 comes
\cite{BGGS-AB-99}
from the up-down asymmetry
of multi-GeV muon-like events measured in the Super-Kamiokande experiment
\cite{SK-atm}.
The values of
$|U_{e4}|^2_{\mathrm{max}}$ and $|U_{\mu4}|^2_{\mathrm{max}}$
as functions of
$|\Delta{m}^2_{41}|$
are shown, respectively, in Figs.~\ref{uel4} and \ref{umu4}.

In the 3+1 schemes we have
\begin{equation}
A_{{\mu}e}
=
4 |U_{\mu4}|^2 |U_{e4}|^2
\,,
\label{01}
\end{equation}
and the bounds (\ref{24}) imply that
\cite{BGG-AB-96}
\begin{equation}
A_{{\mu}e}
\leq
4 |U_{\mu4}|^2_{\mathrm{max}} |U_{e4}|^2_{\mathrm{max}}
\,.
\label{02}
\end{equation}
As one can see from Fig.~\ref{amuel-allowed},
from the LSND region constrained by
the exclusion curves of KARMEN and BNL-E776
and by the bound (\ref{02}),
there are four allowed regions for $A_{{\mu}e}$,
R1, R2, R3, R4 in Eq.~(\ref{Regions}),
where we have
\begin{equation}
A_{{\mu}e}^{\mathrm{min}}
\leq
4 |U_{\mu4}|^2 |U_{e4}|^2
\leq
A_{{\mu}e}^{\mathrm{max}}
\,.
\label{03}
\end{equation}
These regions could be explored in the near future by
the MiniBooNE experiment \cite{MiniBooNE}.
Region R4 is at the limit
of the final NOMAD sensitivity in the
$\nu_\mu\to\nu_e$ channel
\cite{NOMAD-status-99}.

From the lower bound in Eq.~(\ref{03})
and the bounds (\ref{24}),
we obtain
\begin{eqnarray}
&&
|U_{e4}|^2
\geq
\frac{A_{{\mu}e}^{\mathrm{min}}}{4 |U_{\mu4}|^2}
\geq
\frac{A_{{\mu}e}^{\mathrm{min}}}{4 |U_{\mu4}|^2_{\mathrm{max}}}
\,,
\label{04}
\\
&&
|U_{\mu4}|^2
\geq
\frac{A_{{\mu}e}^{\mathrm{min}}}{4 |U_{e4}|^2}
\geq
\frac{A_{{\mu}e}^{\mathrm{min}}}{4 |U_{e4}|^2_{\mathrm{max}}}
\,.
\label{05}
\end{eqnarray}
Therefore, we have
\begin{eqnarray}
&&
\frac{A_{{\mu}e}^{\mathrm{min}}}{4 |U_{\mu4}|^2_{\mathrm{max}}}
\leq
|U_{e4}|^2
\leq
|U_{e4}|^2_{\mathrm{max}}
\,,
\label{06}
\\
&&
\frac{A_{{\mu}e}^{\mathrm{min}}}{4 |U_{e4}|^2_{\mathrm{max}}}
\leq
|U_{\mu4}|^2
\leq
|U_{\mu4}|^2_{\mathrm{max}}
\,.
\label{07}
\end{eqnarray}
Since these bounds
have been obtained without any assumption
on the value of $|U_{s4}|^2$,
they are generally valid in any 3+1 scheme.
The corresponding four allowed regions for
$|U_{e4}|^2$ and $|U_{\mu4}|^2$
are shown,
respectively,
in Figs.~\ref{uel4} and \ref{umu4}.

The lower bounds for $|U_{e4}|^2$ and $|U_{\mu4}|^2$
in Eqs.~(\ref{04}) and (\ref{05})
imply lower bounds for the oscillation amplitudes
$ B_e = 4 |U_{e4}|^2 \left( 1 - |U_{e4}|^2 \right) $
and
$ B_\mu = 4 |U_{\mu4}|^2 \left( 1 - |U_{\mu4}|^2 \right) $.
Let us derive these bounds.

Since $|U_{e4}|^2_{\mathrm{max}}$ is always smaller than 1/2,
the lower bound for $B_e$ is
\begin{equation}
B_e^{\mathrm{min}}
=
\frac{A_{{\mu}e}^{\mathrm{min}}}{|U_{\mu4}|^2_{\mathrm{max}}}
\left( 1 - \frac{A_{{\mu}e}^{\mathrm{min}}}{4 |U_{\mu4}|^2_{\mathrm{max}}} \right)
\,.
\label{08}
\end{equation}
On the other hand,
since $|U_{\mu4}|^2_{\mathrm{max}}$ can be bigger than 1/2,
the lower bound for $B_\mu$ is
\begin{equation}
B_\mu^{\mathrm{min}}
=
\min\left[
\frac{A_{{\mu}e}^{\mathrm{min}}}{|U_{e4}|^2_{\mathrm{max}}}
\left( 1 - \frac{A_{{\mu}e}^{\mathrm{min}}}{4 |U_{e4}|^2_{\mathrm{max}}} \right)
\, , \,
|U_{\mu4}|^2_{\mathrm{max}} \left( 1 - |U_{\mu4}|^2_{\mathrm{max}} \right)
\right]
\,.
\label{09}
\end{equation}

\begin{figure}[p!]
\begin{center}
\includegraphics[bb=89 331 540 763,width=\textwidth]{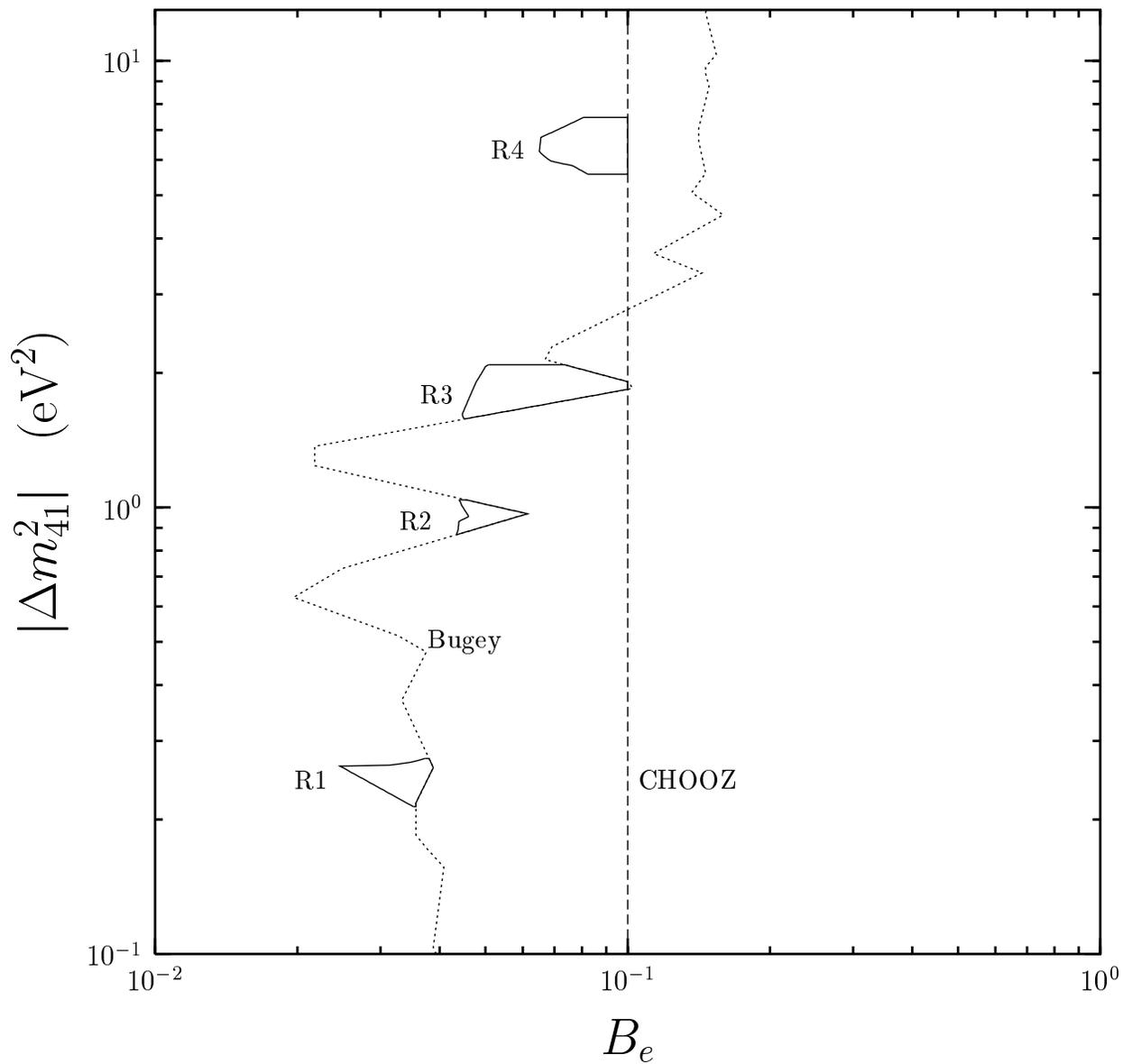}
\end{center}
\caption{ \label{bel}
\textbf{Solid Line:}
Allowed regions.
\textbf{Dotted Line:}
Bugey exclusion curve at 90\% CL \protect\cite{Bugey}.
\textbf{Dashed Line:}
CHOOZ exclusion curve at 90\% CL \protect\cite{CHOOZ}.
}
\end{figure}

\begin{figure}[p!]
\begin{center}
\includegraphics[bb=89 331 540 763,width=\textwidth]{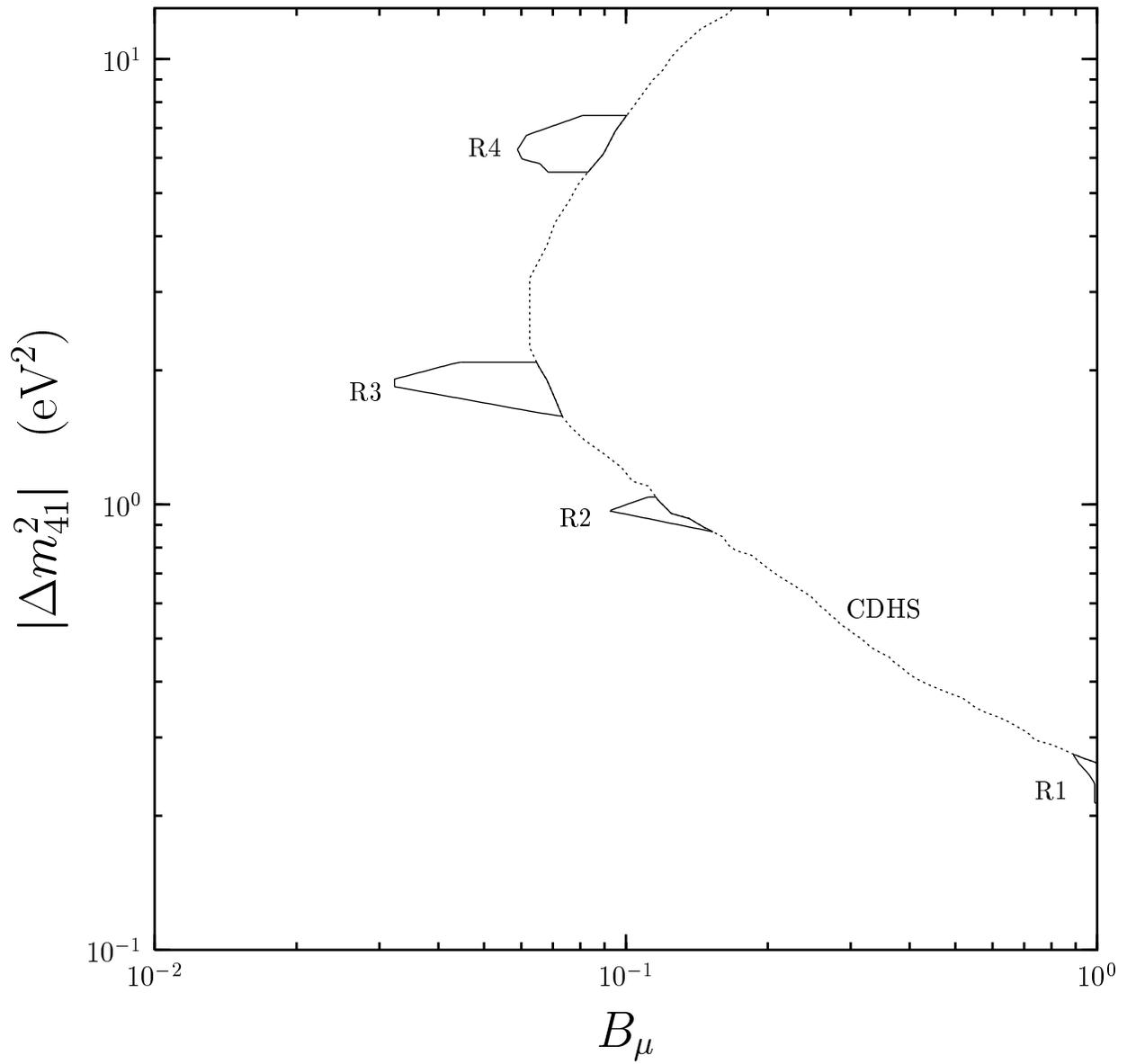}
\end{center}
\caption{ \label{bmu}
\textbf{Solid Line:}
Allowed regions.
\textbf{Dotted Line:}
CDHS exclusion curve at 90\% CL \protect\cite{CDHS}.
}
\end{figure}

Figure \ref{bel} and \ref{bmu}
show the allowed regions
in the
$B_e$--$|\Delta{m}^2_{41}|$
and
$B_\mu$--$|\Delta{m}^2_{41}|$
planes given by the bounds
\begin{equation}
B_\alpha^{\mathrm{min}}
\leq
B_\alpha
\leq
B_\alpha^{\mathrm{max}}
\qquad
(\alpha=e,\mu)
\,.
\label{31}
\end{equation}
One can see that these regions lie just on the left of
the Bugey+CHOOZ (Fig.~\ref{bel})
and CDHS (Fig.~\ref{bmu})
exclusion curves and could be observed in the near future
\cite{Mikaelyan-99-00}.
Let us emphasize that these regions
are generally predicted in any 3+1 scheme,
since they have been obtained independently on any assumption
on the mixing
(as Eq.~(\ref{12}) or Eq.~(\ref{13})).

\section{Large
$\lowercase{\nu_e\to\nu_\tau}$
and
$\lowercase{\nu_\mu\to\nu_\tau}$
short-baseline transitions}
\label{large}

We have now all the elements to calculate the bounds
(\ref{21}) and (\ref{22})
on the amplitudes of short-baseline
$\nu_e\to\nu_\tau$
and
$\nu_\mu\to\nu_\tau$
oscillations
that follow from the assumption (\ref{13}),
$|U_{s4}|^2\ll1$.

\begin{figure}[p!]
\begin{center}
\includegraphics[bb=89 331 540 763,width=\textwidth]{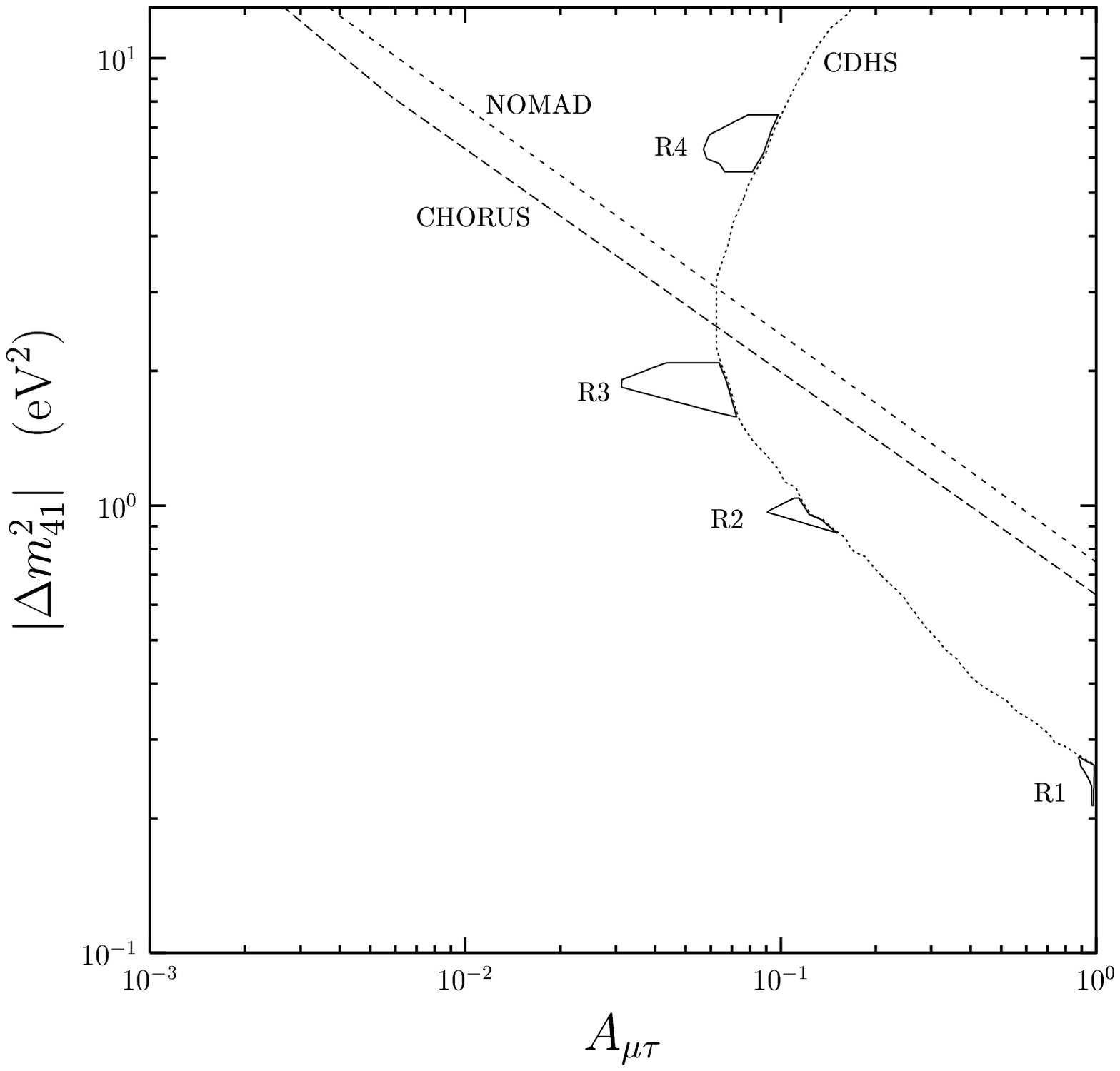}
\end{center}
\caption{ \label{amuta}
\textbf{Solid Line:}
Allowed regions.
\textbf{Long Dashed Line:}
CHORUS exclusion curve at 90\% CL \protect\cite{CHORUS}.
\textbf{Short Dashed Line:}
NOMAD exclusion curve at 90\% CL \protect\cite{NOMAD}.
\textbf{Dotted Line:}
CDHS exclusion curve at 90\% CL \protect\cite{CDHS}.
}
\end{figure}

Figure \ref{amuta}
shows the allowed regions
in the
$A_{\mu\tau}$--$|\Delta{m}^2_{41}|$
plane given by the bounds
(\ref{22}).
One can see that the region R4
is excluded by the negative results
of the CHORUS \cite{CHORUS} and NOMAD \cite{NOMAD} experiments.
The other three regions are possible
and predict relatively large
oscillation amplitudes that could be observed
in the near future,
especially the two regions R2 and R3
in which
$A_{\mu\tau} \sim 4 \times 10^{-2} - 10^{-1}$.

\begin{figure}[p!]
\begin{center}
\includegraphics[bb=89 331 540 763,width=\textwidth]{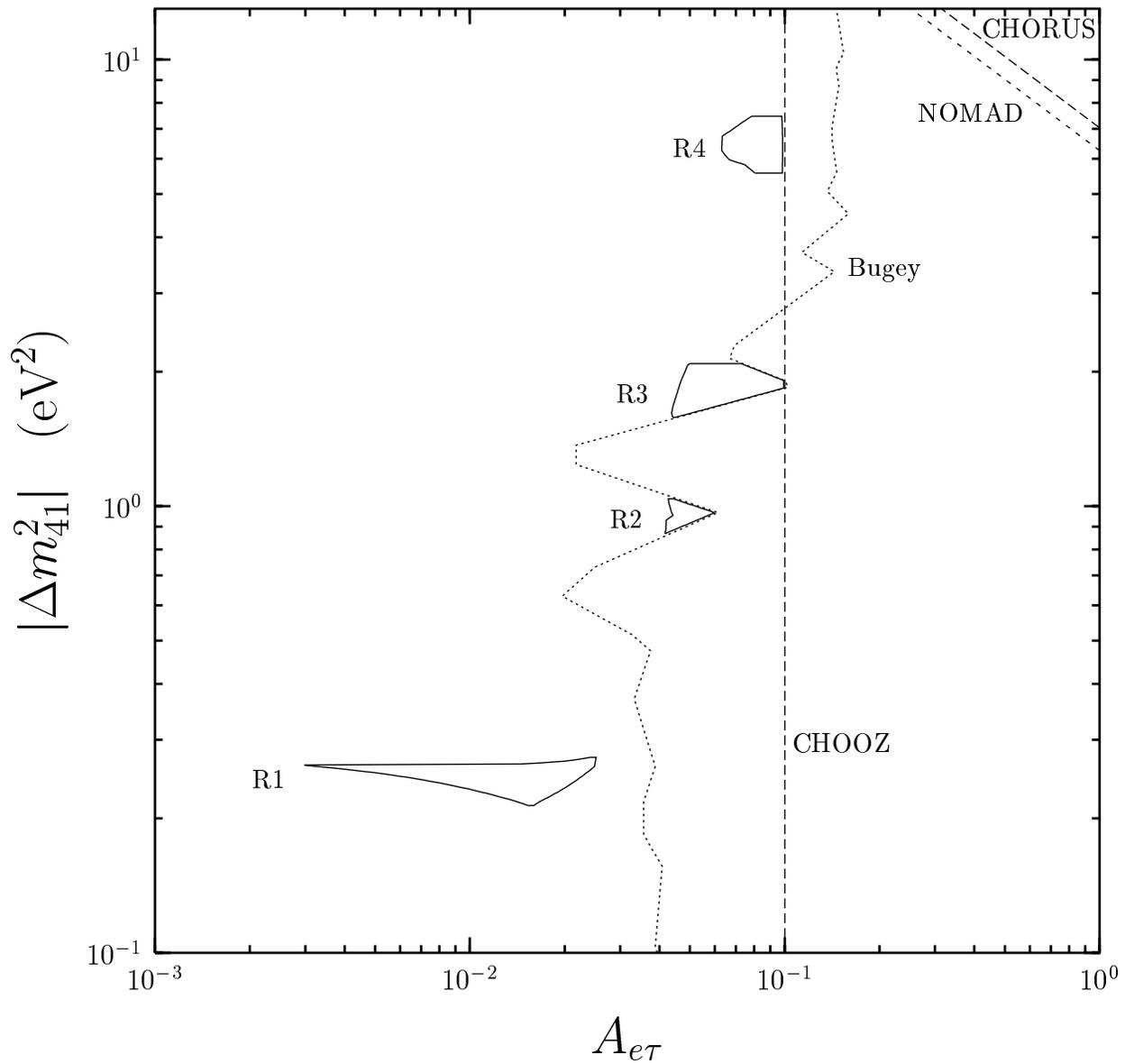}
\end{center}
\caption{ \label{aelta}
\textbf{Solid Line:}
Allowed regions.
\textbf{Long Dashed Line:}
CHORUS exclusion curve at 90\% CL \protect\cite{CHORUS}.
\textbf{Short Dashed Line:}
NOMAD exclusion curve at 90\% CL \protect\cite{NOMAD}.
\textbf{Dotted Line:}
Bugey exclusion curve at 90\% CL \protect\cite{Bugey}.
\textbf{Dashed Line:}
CHOOZ exclusion curve at 90\% CL \protect\cite{CHOOZ}.
}
\end{figure}

Figure \ref{aelta}
shows the allowed regions
in the
$A_{e\tau}$--$|\Delta{m}^2_{41}|$
plane given by the bounds
(\ref{21}).
One can see that these regions
predict relatively large
oscillation amplitudes,
but unfortunately
lie rather far from the CHORUS and NOMAD exclusion curves
(except the region R4
that is excluded by Fig.~\ref{amuta}).
Therefore,
even under the favorable assumption
(\ref{13}),
it will be very difficult to observe
$\nu_e\to\nu_\tau$
transitions in short baseline experiments
with conventional neutrino beams,
but large transitions
could be observed with a $\nu_e$
beam from a neutrino factory
\cite{nufactory}.

\begin{figure}[p!]
\begin{center}
\includegraphics[bb=89 331 540 763,width=\textwidth]{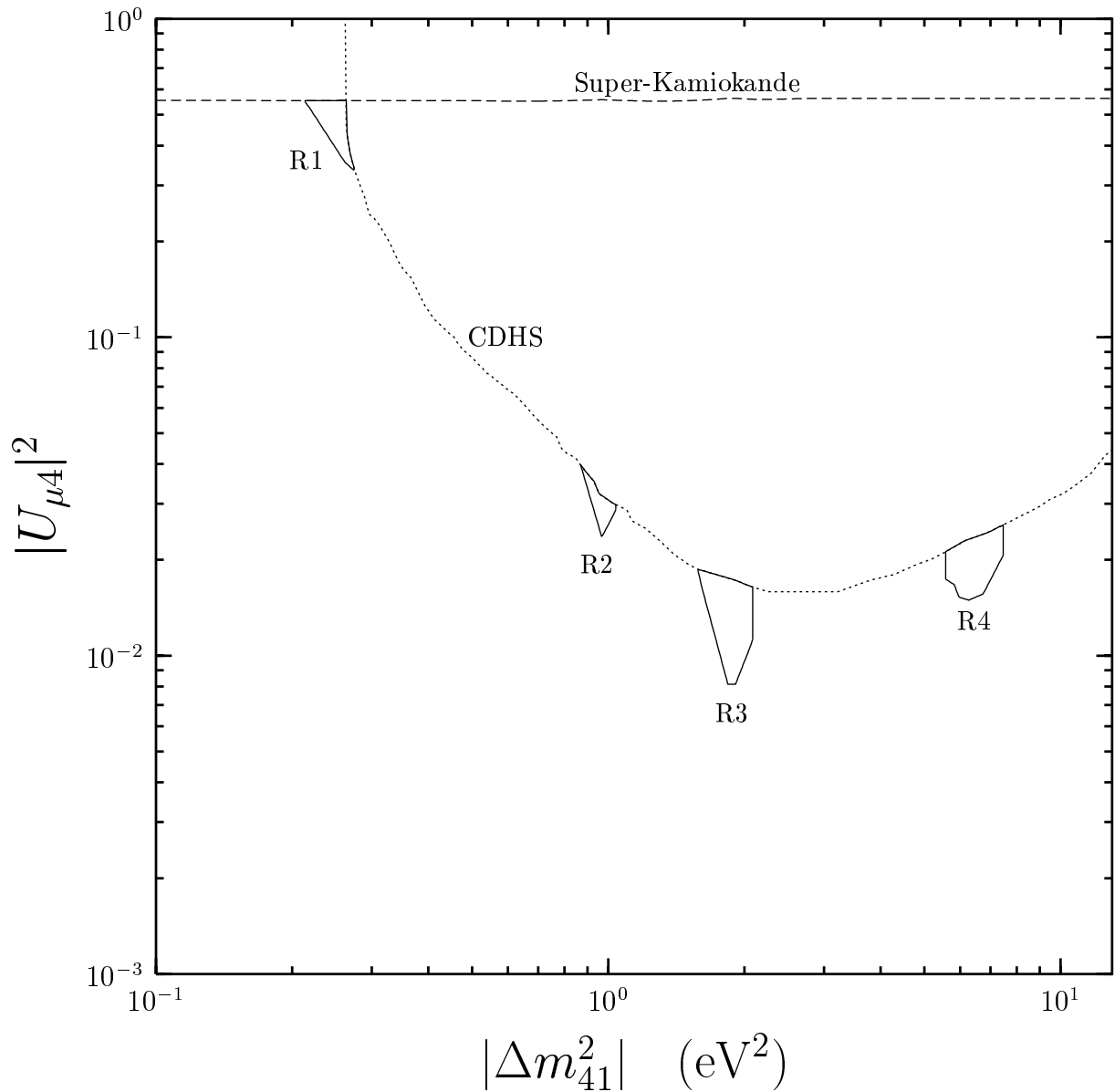}
\end{center}
\caption{ \label{umu4}
\textbf{Dotted Line:}
$|U_{\mu4}|^2_{\mathrm{max}}$
obtained from
the exclusion curve of the
CDHS accelerator $\nu_\mu$ disappearance experiment
\protect\cite{CDHS}
(Eq.~(\ref{26})).
\textbf{Dashed Line:}
$|U_{\mu4}|^2_{\mathrm{max}}$
obtained from
the up-down asymmetry
of multi-GeV muon-like events
measured in the Super-Kamiokande experiment
\protect\cite{SK-atm}
(Eq.~(\ref{26})).
\textbf{Solid Line:}
Allowed regions (Eq.~(\ref{07})).
}
\end{figure}

As one can see from Fig.~\ref{uel4},
the bound in Eq.~(\ref{24})
imply that $|U_{e4}|^2$
is very small in all the three allowed regions
in Fig.~\ref{amuta}.
On the other hand,
one can see from Fig.~\ref{umu4} that
$|U_{\mu4}|^2$
is small in the two allowed regions R2
and
R3,
but it is large in the region R1.
As a consequence of the unitarity of the mixing matrix
and the assumption (\ref{13}),
we have
$1-|U_{\tau4}|^2 \ll 1$
in R2 and R3,
whereas
$|U_{\tau4}|^2$
can be as small as about 1/2
in R1.
The prediction for solar,
atmospheric and long-baseline experiments
depend on the value of
$|U_{\mu4}|^2$.
There are two possibilities:

\begin{enumerate}

\item
$|U_{\mu4}|^2 \ll 1$ in regions R2 and R3
in Fig.~\ref{umu4}.
In this case
$1-|U_{\tau4}|^2 \ll 1$
and
the atmospheric neutrino anomaly is due to dominant
$\nu_\mu\to\nu_s$
transitions.
This possibility is disfavored by present
Super-Kamiokande and MACRO data
\cite{SK-atm,MACRO},
but still it is not completely excluded
(see
\cite{Concha-atm-00,Foot-SK-00,Fogli-Lisi-Marrone-fouratm-00}).
Obviously,
in this case long-baseline
$\nu_\mu\to\nu_\tau$ transitions
are suppressed with respect to the dominant
$\nu_\mu\to\nu_s$ transitions,
that are almost entirely responsible
of the disappearance of $\nu_\mu$'s.

The solar neutrino problem is due to an
approximately equal mixture of
$\nu_e\to\nu_\mu$ and $\nu_e\to\nu_s$ transitions,
which is allowed by the data.
This has been shown in
Refs.~\cite{Concha-foursolar-00,Concha-foursolar-update}
in the framework of the 2+2 schemes in Fig.~\ref{4schemes}.
However,
the results obtained in
Refs.~\cite{Concha-foursolar-00,Concha-foursolar-update}
are valid also in the 3+1 schemes,
because the formalism of solar neutrino oscillations
in 3+1 schemes is identical to that in 2+2 schemes
\cite{DGKK-99,Concha-foursolar-00,Concha-foursolar-update}.
Indeed,
from Eqs.~(\ref{24}) and (\ref{25})
we already know that
the results of the Bugey and CHOOZ experiment imply that
$|U_{e4}|^2$ is very small.
Furthermore,
taking into account the hierarchy
\begin{equation}
\Delta{m}^2_{21} = \Delta{m}^2_{\mathrm{SUN}}
\ll
|\Delta{m}^2_{31}| = \Delta{m}^2_{\mathrm{ATM}}
\ll
|\Delta{m}^2_{41}| = \Delta{m}^2_{\mathrm{SBL}}
\,,
\label{hierarchy}
\end{equation}
the effective survival probability of $\bar\nu_e$ and $\nu_e$
in long-baseline experiments
is given by
\begin{equation}
P_{\nu_e\to\nu_e}^{\mathrm{LBL}}
=
1
-
4 |U_{e3}|^2 \left( 1 - |U_{e3}|^2 \right)
\sin^2\left(\frac{\Delta{m}^2_{31}L}{4E}\right)
+
|U_{e4}|^4
\,,
\label{LBL-ee}
\end{equation}
where $L$ is the propagation distance
and $E$ is the neutrino energy.
Neglecting $|U_{e4}|^4$,
Eq.~(\ref{LBL-ee})
has the same structure as the usual two-generation survival
probability
(see \cite{BGG-review-98}),
with $\sin^2 2\vartheta$
replaced by
$4 |U_{e3}|^2 \left( 1 - |U_{e3}|^2 \right)$
and
$\Delta{m}^2$
replaced by
$\Delta{m}^2_{31}$.
The bound on $\sin^2 2\vartheta$
obtained in the CHOOZ experiment \cite{CHOOZ}
for
$|\Delta{m}^2_{31}| \gtrsim 10^{-3} \, \mathrm{eV}^2$
implies that
$|U_{e3}|^2 \lesssim 2.6 \times 10^{-2}$.
Therefore,
both
$|U_{e3}|^2$
and
$|U_{e4}|^2$
are very small,
\begin{equation}
|U_{e3}|^2 \lesssim 3 \times 10^{-2}
\,
\qquad
|U_{e4}|^2 \lesssim 3 \times 10^{-2}
\,,
\label{Ue3}
\end{equation}
and can be neglected in
the study of solar neutrino oscillations,
as done in
Refs.~\cite{DGKK-99,Concha-foursolar-00,Concha-foursolar-update}.
In other words,
the production and detection of the mass eigenstates
$\nu_3$ and $\nu_4$
is negligibly small in solar neutrino experiments.
Moreover, because of the hierarchy (\ref{hierarchy}),
there are no matter-induced transitions
from $\nu_1$, $\nu_2$
to $\nu_3$, $\nu_4$.
Hence,
$\nu_3$ and $\nu_4$
effectively decouple in solar neutrino oscillations,
leading to the formalism
described
in Refs.~\cite{DGKK-99,%
Concha-foursolar-00,Concha-foursolar-update}.

\item
$|U_{\mu4}|^2 \simeq 0.33 - 0.55$ in region R1
in Fig.~\ref{umu4}.
In this case
the solar neutrino problem is due to a mixture of
$\nu_e\to\nu_\mu$, $\nu_e\to\nu_\tau$
and $\nu_e\to\nu_s$ transitions,
that is allowed by data as in the previous case.

The atmospheric neutrino anomaly is due a mixture of
$\nu_\mu\to\nu_\tau$ and $\nu_\mu\to\nu_s$
transitions.
In Refs.~\cite{Yasuda-fouratm-00,%
Lisi-Nu2000,%
Fogli-Lisi-Marrone-fouratm-00}
it has been shown that
a mixture of
$\nu_\mu\to\nu_\tau$ and $\nu_\mu\to\nu_s$
transitions
in the framework of 2+2 schemes
is allowed by the atmospheric neutrino data.
This indicates that
such a mixture should be allowed also
in the framework of 3+1 schemes.
The groups specialized in
the analysis of atmospheric neutrino data
could check this possibility using the formalism presented
in Appendix~\ref{append}.
The existence of mixed
$\nu_\mu\to\nu_\tau$ and $\nu_\mu\to\nu_s$
transitions
can also be checked by comparing the rates
of $\nu_\mu$ disappearance and $\nu_\mu\to\nu_\tau$
appearance in future long-baseline experiments
with $\nu_\mu$ beams.
The formalism that allows to describe
these oscillation channels
in the framework of 3+1 four-neutrino schemes
is presented
in Appendix~\ref{append}.

\end{enumerate}

These predictions are testable
in future experiments,
especially measuring the
percentage of
transitions into active and sterile neutrinos
in solar, atmospheric and long-baseline experiments
(that should measure the same transitions
observed in atmospheric neutrino experiments).
Taking into account also the prediction of large
$\nu_\mu\to\nu_\tau$
and
$\nu_e\to\nu_\tau$
in short-baseline experiments,
the schemes under consideration can be checked and
possibly falsified in the near future.

\section{Conclusions}
\label{Conclusions}

In conclusion,
we have considered the four-neutrino 3+1 schemes
in Fig.~\ref{4schemes},
that are marginally allowed by present data
(see Fig.~\ref{amuel-allowed}).
We have identified four small regions
in the parameter space of neutrino mixing
compatible with all data
and
we have derived general upper and lower bounds
for the elements
$U_{e4}$ and $U_{\mu4}$
of the mixing matrix
and on the oscillation amplitudes in
short-baseline
$\bar\nu_e$ and $\nu_\mu$
disappearance experiments.
The corresponding
$\nu_\mu\to\nu_e$ transitions
and
$\nu_e$ and $\nu_\mu$
disappearance in short-baseline experiments
are relatively large
and could be observed in future dedicated experiments.
Assuming a small mixing of the sterile neutrino
with the isolated mass eigenstate,
$|U_{s4}|^2 \ll 1$,
we have shown that
large
$\nu_\mu\to\nu_\tau$
and
$\nu_e\to\nu_\tau$
transitions
are predicted in short-baseline experiments.
We have also discussed the implications
of $|U_{s4}|^2 \ll 1$
for solar, atmospheric
and long-baseline neutrino oscillation experiments
and we have presented
in Appendix~\ref{append}
the formalism describing
in general 3+1 schemes
the oscillations in matter
of atmospheric neutrinos and
neutrinos
in long-baseline $\nu_\mu$ disappearance and
$\nu_\mu\to\nu_\tau$ appearance
experiments.
Finally,
let us remark that
the four 3+1 schemes in Fig.~\ref{4schemes}
are equivalent for solar, atmospheric
and short-baseline
neutrino oscillation experiments,
but they may be distinguishable
in long-baseline
$\nu_\mu\to\nu_e$
experiments
if $|U_{e3}|^2$
is not too small \cite{DGKK-preparation},
or
through their different effects
in tritium $\beta$ decay
and neutrinoless double-$\beta$ decay
(see \cite{Bilenky-Petcov-RMP-87,CWKim-book,BGG-review-98})
and through neutrino oscillations
in supernovae
(see
\cite{Peltoniemi-supernovae-sterile,Fuller-supernovae-sterile})
and in the early universe
(see
\cite{Foot-Volkas-BBN-sterile,Kirilova-Chizhov-BBN,%
Dolgov-BBN-sterile,%
Esposito-BBN-sterile,Shi-Fuller-BBN}).

\acknowledgments 
C.G. would like to thank
S. Bilenky,
E. Lisi,
O. Peres,
F. Ronga,
and
A. Smirnov
for useful discussions during the
NOW2000 workshop.
C.G. would also like to express his gratitude to
the Korea Institute for Advanced Study (KIAS)
for warm hospitality during the completion of this work.

\newpage 
\appendix
 
\section{Formalism of atmospheric neutrino oscillations} 
\label{append}

In this appendix we derive in a concise way
the formalism
in 3+1 schemes of atmospheric neutrino oscillations
and oscillations in
long-baseline
$\nu_\mu\to\nu_\tau$
and
$\nu_\mu$
disappearance
experiments.
We take into account matter effects,
following a method similar to that used in Ref.~\cite{DGKK-99}
in the case of 2+2 schemes.
As explained in the main text,
the formalism of solar neutrino oscillations
in 3+1 schemes is identical to that in 2+2 schemes
\cite{DGKK-99,Concha-foursolar-00,Concha-foursolar-update}.

From Eq.~(\ref{Ue3})
we know that $U_{e3}$ and $U_{e4}$
can be neglected.
Choosing the ordering
$\nu_s$,
$\nu_e$,
$\nu_\mu$,
$\nu_\tau$
for the flavor neutrino fields,
the mixing matrix can be written as
\begin{equation}
U = V_{34} \, V_{14} \, V_{13} \, V_{12}
\,,
\label{U-atm}
\end{equation}
where
\begin{equation}
(V_{ij})_{ab}
=
\delta_{ab}
+
\left( \cos{\vartheta_{ij}} - 1 \right)
\left( \delta_{ia} \delta_{ib} + \delta_{ja} \delta_{jb} \right)
+
\sin{\vartheta_{ij}}
\left( \delta_{ia} \delta_{jb} - \delta_{ja} \delta_{ib} \right)
\label{atm-rotation}
\end{equation}
represents a rotation by an angle $\vartheta_{ij}$
in the $i$-$j$ plane.
We have also neglected possible
CP-violating phases.
The evolution of the neutrino flavor amplitudes
$\psi_\alpha$ ($\alpha=s,e,\mu,\tau$)
in vacuum and in matter is given by the
MSW equation
\cite{MSW}
\begin{equation}
i \, \frac{ \mathrm{d}}{ \mathrm{d} x } \, \Psi
=
\mathcal{H} \, \Psi
\,,
\label{evolution1}
\end{equation}
where
$
\Psi
=
\left(
\psi_s,
\psi_e,
\psi_\mu,
\psi_\tau
\right)^T
$
and
$\mathcal{H}$ is the effective Hamiltonian
\begin{equation}
\mathcal{H}
=
\frac{1}{2p}
\left( U \, \mathcal{M}_0^2 \, U^\dagger + \mathcal{A} \right)
\,.
\label{hamiltonian1}
\end{equation}
Here $p$ is the neutrino momentum,
\begin{equation}
\mathcal{M}_0^2
=
\mathrm{diag}(0,0,\Delta{m}^2_{31},\Delta{m}^2_{41})
\label{mass1}
\end{equation}
is the mass-squared matrix,
in which we neglected $\Delta{m}^2_{21}$
according to the hierarchical relation (\ref{hierarchy}),
and $\mathcal{A}$ is the matrix
\begin{equation}
\mathcal{A} = \mathrm{diag}(-A_{NC},A_{CC},0,0)
\,,
\label{A}
\end{equation}
with
$ A_{CC} = 2 p V_{CC} $
and
$ A_{NC} = 2 p V_{NC} $,
where
$V_{CC}$
and
$V_{NC}$
are,
respectively,
the matter-induced
charged-current and neutral-current potentials
(see \cite{Mikheev-Smirnov-uspekhi-87,Bilenky-Petcov-RMP-87,%
Kuo-Pantaleone-RMP-89,CWKim-book,BGG-review-98}).
For atmospheric neutrinos
propagating in the Earth
and accelerator neutrinos in long-baseline experiments
$
|A_{CC}| \sim |A_{NC}| \sim |\Delta{m}^2_{31}|
\ll 
|\Delta{m}^2_{41}|
$.
In Eq.~(\ref{hamiltonian1})
we have neglected a common phase for the flavor amplitudes,
that is irrelevant for neutrino oscillations.

\begin{figure}[p!]
\begin{center}
\includegraphics[bb=13 668 522 827,width=\textwidth]{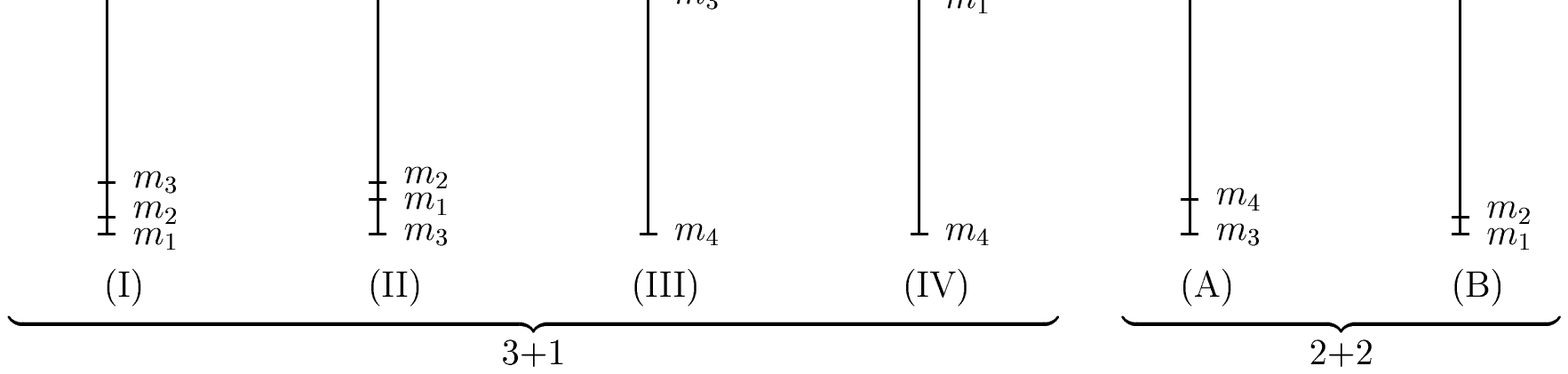}
\end{center}
\caption{ \label{4schemes}
Qualitative illustration of the possible four-neutrino schemes.
}
\end{figure}

The evolution equation (\ref{evolution1})
is most easily solved in the rotated basis
$ \Psi' = V_{14}^T V_{34}^T \Psi
=
\left( \psi'_1, \psi'_2, \psi'_3, \psi'_4 \right)^T
$
that obeys to a similar evolution equation,
with
$\mathcal{H}$
replaced by
\begin{equation}
\mathcal{H}'
=
\frac{1}{2p}
\left(
V_{13} \mathcal{M}_0^2 V_{13}^T
+
V_{14}^T \mathcal{A} V_{14}
\right)
\,
\label{Hp}
\end{equation}
where we have taken into account the fact that
$ V_{12} \mathcal{M}_0^2 V_{12}^T = \mathcal{M}_0^2 $
and
$ V_{34}^T \mathcal{A} V_{34} = \mathcal{A} $,
which lead to a significant
simplification of the evolution equation.
The explicit form of
$\mathcal{H}'$
is
\begin{equation}
\mathcal{H}'
=
\frac{1}{4p}
\left(
\begin{array}{cccc}
\scriptstyle
( 1 - c_{2\vartheta_{13}} ) \Delta{m}^2_{31}
- 2 c^2_{\vartheta_{14}} A_{NC}
&
\scriptstyle
0
&
\scriptstyle
s_{2\vartheta_{13}} \Delta{m}^2_{31}
&
\scriptstyle
- s_{2\vartheta_{14}} A_{NC}
\\
\scriptstyle
0
&
\scriptstyle
2 A_{CC}
&
\scriptstyle
0
&
\scriptstyle
0
\\
\scriptstyle
s_{2\vartheta_{13}} \Delta{m}^2_{31}
&
\scriptstyle
0
&
\scriptstyle
( 1 + c_{2\vartheta_{13}} ) \Delta{m}^2_{31}
&
\scriptstyle
0
\\
\scriptstyle
- s_{2\vartheta_{14}} A_{NC}
&
\scriptstyle
0
&
\scriptstyle
0
&
\scriptstyle
2 \Delta{m}^2_{41} - 2 s^2_{\vartheta_{14}} A_{NC}
\end{array}
\right)
\,,
\label{Hpe}
\end{equation}
with
$c_{\vartheta} \equiv \cos\vartheta$
and
$s_{\vartheta} \equiv \sin\vartheta$.
This equation shows that the amplitudes
$\psi'_2$
and
$\psi'_4$
evolve independently
(remember that $|\Delta{m}^2_{41}| \gg |A_{NC}|$),
with phases given by the energy eigenvalues
\begin{equation}
E'_2
=
A_{CC} / 2p
=
V_{CC}
\,,
\quad
E'_4 \simeq \Delta{m}^2_{41} / 2 p
\,,
\label{Ep14}
\end{equation}
whereas the evolutions of the amplitudes
$\psi'_1$
and
$\psi'_3$
are coupled and
there is a resonance in the 1-3 sector for
\begin{equation}
- \cos^2{\vartheta_{14}} \, A_{NC}
=
\cos{2\vartheta_{13}} \, \Delta{m}^2_{31}
\,.
\label{Res}
\end{equation}
Notice that for neutrinos the left-hand side
of Eq.~(\ref{Res}) is positive,
because $A_{NC} \leq 0$,
whereas for antineutrinos it is negative
because
$A_{NC}$
must be replaced by $\overline{A}_{NC} = - A_{NC}$.
Hence,
if
$\Delta{m}^2_{31} > 0$
(schemes I and IV in Fig.~\ref{4schemes})
and
$\vartheta_{13} < \pi/4$
or
$\Delta{m}^2_{31} < 0$
(schemes II and III)
and
$\vartheta_{13} > \pi/4$,
there can be a resonance for neutrinos,
otherwise the resonance can be for antineutrinos.

One can calculate the evolution
of the amplitudes
$\psi'_1$
and
$\psi'_3$
solving numerically the two coupled equations
generated by the 1-3 sector of the effective Hamiltonian
$\mathcal{H}'$.
Another common method
for the solution of the evolution equation
is to divide the Earth
interior into shell with constant density,
calculate the evolution of the amplitudes
$\Psi'$
in each shell and match the amplitudes
in the flavor basis at the shell boundaries.
In this case,
in each shell
the amplitudes
$\psi'_1$
and
$\psi'_3$
can be written as
\begin{equation}
\psi'_1
=
\cos\vartheta_{13}^M \psi^M_1
+
\sin\vartheta_{13}^M \psi^M_3
\,,
\quad
\psi'_3
=
- \sin\vartheta_{13}^M \psi^M_1
+ \cos\vartheta_{13}^M \psi^M_3
\,,
\label{psiM}
\end{equation}
where
$\psi^M_1$ and $\psi^M_3$
are the amplitudes of the energy eigenstates
that evolve with phases given by the energy eigenvalues
\begin{equation}
E^M_{1,3}
=
\frac{1}{4p}
\left[
\Delta{m}^2_{31}
- 
\cos^2\vartheta_{14} A_{NC}
\mp
\sqrt{
\left(
\cos2\vartheta_{13} \Delta{m}^2_{31}
+
\cos^2\vartheta_{14} \, A_{NC}
\right)^2
+
\left( \sin2\vartheta_{13} \Delta{m}^2_{31} \right)^2
}
\right]
\,.
\label{EP23}
\end{equation}
The effective mixing angle in matter
$\vartheta_{13}^M$
is given by
\begin{equation}
\tan2\vartheta_{13}^M
=
\frac{ \sin2\vartheta_{13} \Delta{m}^2_{31} }
{ \cos2\vartheta_{13} \Delta{m}^2_{31}
+ \cos^2\vartheta_{14} \, A_{NC} }
\,,
\label{p23m}
\end{equation}
from which one can see that
$\vartheta_{13}^M = \pi/4$
when the resonance condition (\ref{Res})
is satisfied,
and
$\vartheta_{13}^M \to \pi/2$
in a very dense medium,
where
$
|\cos^2\vartheta_{14} A_{NC}|
\gg
|\cos2\vartheta_{13} \Delta{m}^2_{31}|
$
if
$
\cos2\vartheta_{13} \Delta{m}^2_{31} > 0
$,
whereas
$\vartheta_{13}^M \to 0$
in a very dense medium
if
$
\cos2\vartheta_{13} \Delta{m}^2_{31} < 0
$.

The connection between the
amplitudes
$\Psi'$
and the flavor amplitudes
$\Psi$,
needed for the calculation of the probability
of flavor transitions,
is given by
\begin{eqnarray}
&&
\psi_s
=
\cos\vartheta_{14} \psi'_1
+
\sin\vartheta_{14} \psi'_4
\,,
\nonumber
\\
&&
\psi_e
=
\psi'_2
\,,
\nonumber
\\
&&
\psi_\mu
=
-
\sin\vartheta_{14} \sin\vartheta_{34} \psi'_1
+
\cos\vartheta_{34} \psi'_3
+
\cos\vartheta_{14} \sin\vartheta_{34} \psi'_4
\,,
\nonumber
\\
&&
\psi_\tau
=
-
\sin\vartheta_{14} \cos\vartheta_{34} \psi'_1
-
\sin\vartheta_{34} \psi'_3
+
\cos\vartheta_{14} \cos\vartheta_{34} \psi'_4
\,.
\label{connection}
\end{eqnarray}
The second line
of Eq.~(\ref{connection})
imply that electron neutrinos do not oscillate
in atmospheric neutrino experiments
(remember that $\psi'_2$ evolves independently).
This is due to the approximation
$U_{e3} = 0$,
motivated by Eq.~(\ref{Ue3}),
which also imply that
the charged-current matter potential $V_{CC}$,
felt only by $\nu_e$,
is irrelevant for atmospheric neutrino oscillations
(only the neutral-current potential
$V_{NC}$,
felt by $\nu_\mu$ and $\nu_\tau$,
enter in Eqs.~(\ref{Res})--(\ref{p23m})).
On the other hand,
simultaneous
$\nu_\mu\to\nu_\tau$
and
$\nu_\mu\to\nu_s$
transitions
are allowed,
with pure
$\nu_\mu\to\nu_\tau$
transitions in the limit
$\cos\vartheta_{14}=0$,
that corresponds to
$U_{s4}=1$,
and
pure
$\nu_\mu\to\nu_s$
transitions in the limit
$\cos\vartheta_{14}\cos\vartheta_{34}=1$,
that corresponds to
$U_{\tau4}=1$.

Let us consider finally
long baseline experiments in which the neutrino beam
travels in the crust of the Earth,
where the matter density is practically constant.
The probability of
$\nu_\alpha\to\nu_\beta$
transitions
is given by
\begin{eqnarray}
P_{\nu_\alpha\to\nu_\beta}^{\mathrm{LBL}}
&=&
\left|
U_{\alpha1}^M U_{\beta1}^M
+
U_{\alpha3}^M U_{\beta3}^M
\,
\exp\left( -i \frac{\Delta_{31}^M L}{2 p} \right)
\right|^2
+
|U_{\alpha4}^M|^2
|U_{\beta4}^M|^2
\nonumber
\\
&=&
- 4
U_{\alpha1}^M U_{\beta1}^M
U_{\alpha3}^M U_{\beta3}^M
\sin^2\left( \frac{\Delta_{31}^M L}{4 p} \right)
+
2
|U_{\alpha4}^M|^2
|U_{\beta4}^M|^2
\,,
\label{PLBL}
\end{eqnarray}
where $L$ is the propagation distance,
\begin{equation}
\Delta_{31}^M
=
2p \left( E_3^M - E_1^M \right)
=
\sqrt{
\left(
\cos2\vartheta_{13} \Delta{m}^2_{31}
+
\cos^2\vartheta_{14} \, A_{NC}
\right)^2
+
\left( \sin2\vartheta_{13} \Delta{m}^2_{31} \right)^2
}
\,,
\label{Delta31M}
\end{equation}
and
$U^M$
is the effective mixing matrix in matter,
\begin{equation}
U^M = V_{34} V_{14} V_{13}^M
=
\left(
\begin{array}{cccc} \scriptstyle
c_{\vartheta_{14}} c_{\vartheta_{13}^M}
&
\scriptstyle
0
&
\scriptstyle
c_{\vartheta_{14}} s_{\vartheta_{13}^M}
&
\scriptstyle
s_{\vartheta_{14}}
\\
\scriptstyle
0
&
\scriptstyle
1
&
\scriptstyle
0
&
\scriptstyle
0
\\
\scriptstyle
-
s_{\vartheta_{34}} s_{\vartheta_{14}} c_{\vartheta_{13}^M}
-
c_{\vartheta_{34}} s_{\vartheta_{13}^M}
&
\scriptstyle
0
&
\scriptstyle
-s_{\vartheta_{34}} s_{\vartheta_{14}} s_{\vartheta_{13}^M}
+
c_{\vartheta_{34}} c_{\vartheta_{13}^M}
&
\scriptstyle
s_{\vartheta_{34}} c_{\vartheta_{14}}
\\
\scriptstyle
-
c_{\vartheta_{34}} s_{\vartheta_{14}} c_{\vartheta_{13}^M}
+
s_{\vartheta_{34}} s_{\vartheta_{13}^M}
&
\scriptstyle
0
&
\scriptstyle
-
c_{\vartheta_{34}} s_{\vartheta_{14}} s_{\vartheta_{13}^M}
-
s_{\vartheta_{34}} c_{\vartheta_{13}^M}
&
\scriptstyle
c_{\vartheta_{34}} c_{\vartheta_{14}}
\end{array}
\right)
\,,
\label{UM}
\end{equation}
where $V_{13}^M$
is equal to $V_{13}$
with
$\vartheta_{13}$
replaced by
$\vartheta_{13}^M$.
The expression (\ref{PLBL})
can be used to analyze the data
of long-baseline
$\nu_\mu\to\nu_\tau$
and
$\nu_\mu$
disappearance
experiments
(the analysis of long-baseline $\nu_\mu\to\nu_e$
data requires the relaxation of the approximation
$U_{e3}=0$,
leading to a significant complication of the formalism
\cite{DGKK-preparation}).


\end{document}